\documentclass[twocolumn,twocolappendix]{aastex631}
\begin{document}
\title{Spectral Shapes of Pair Annihilation Line Emission in Magnetar Giant Flares}
\author[0000-0001-6010-714X]{Tomoki Wada}
\affiliation{Frontier Research Institute for Interdisciplinary Sciences, Tohoku University, Sendai, Japan}
\affiliation{Astronomical Institute, Graduate School of Science, Tohoku University, Sendai, Japan}
\affiliation{Department of Physics, National Chung Hsing University, Taichung, Taiwan}
\author[0000-0003-2579-7266]{Shigeo S Kimura}
\affiliation{Frontier Research Institute for Interdisciplinary Sciences, Tohoku University, Sendai, Japan}
\affiliation{Astronomical Institute, Graduate School of Science, Tohoku University, Sendai, Japan}
%
\begin{abstract}
We investigate the gamma-ray spectrum in the MeV range arising from
electron-positron pair annihilation in fireballs associated with 
magnetar giant flares (MGFs), motivated by the recent observation of 
a MeV gamma-ray line feature in a bright gamma-ray burst, GRB~221009A. 
We develop an analytic model of line emission, demonstrating that 
relativistic beaming results in a broadened, power-law spectral feature
with photon index $-1$. 
We then perform Monte Carlo radiative transfer simulations incorporating 
electron-positron pair production, annihilation, and Compton scattering. 
The dependence of the emergent spectrum on the baryon loading is also examined, showing that a baryon-poor fireball is more favorable for the detection of MeV gamma rays.
We further assess the detectability of the line component. The simulation
results indicate that a power-law MeV component from the initial spike
of a Galactic MGFs could be observed with current 
instruments, such as {\it Fermi}/GBM, and will be well within the 
reach of upcoming  MeV gamma-ray satellites, which
are expected to detect ${\mathcal O}(100)$ photons from such events.
\end{abstract}
%
%
\section{Introduction} \label{sec:intro}
%
%
Fireballs are high-temperature, optically thick,
radiation-dominated plasma that emerges in various
high-energy astronomical phenomena. They are  formed when
a large amount of energy is deposited into a compact
region, causing the plasma to reach high temperature
where electron-positron pair production is ignited.
Some of high-energy emission in astronomical transients
are thought to originate from these fireballs. Part of 
the emission associated with gamma-ray bursts (GRBs; e.g.,
\citealp{Goo1986,Pac1986,ShePir1990,MesLag1993,MesRee2000}; 
and see \citealp{Pir1999,Zha2018G} for reviews), magnetar 
giant flares (MGFs; \citealp{ThoDun1995,ThoDun2001,NakPir2005};
and see \citealp{KasBel2017} for reviews), and less energetic,
magnetar short bursts (MSBs) are possible examples of those 
astronomical transients. In the fireballs in MGFs and some of MSBs
\citep{ThoDun2001,Iok2020,YanZha2021,WadIok2023,WadShi2024, WadAsa2025,ShiWad2025},
the initially opaque fireball expands and is accelerated
up to a relativistic speed, undergoing adiabatic cooling. This 
cooling reduces the number density of electron-positron pairs 
in thermal equilibrium, eventually allowing photons to decouple 
and escape from the plasma. 
Consequently, the fireball emits high-energy radiation.

%
%
GRB~221009A, often referred to as the ``Brightest of All Time,"
was discovered on October 9, 2022. This is the brightest GRB
ever observed so far
\citep{AnAnt2023,FreSvi2023,KanAga2023,LesVer2023,LHAASO2023,AxeAje2024,KonWan2024,RavSal2024,ZhaXio2024}.
The associated host galaxy is at
redshift $z\sim 0.15$ \citep{MalLec2023},
indicating an isotropic-equivalent luminosity up to
$\sim 1.5\times10^{55}\,{\rm erg\,s^{-1}}$ \citep{AnAnt2023}.
Partly because of its extreme isotropic-equivalent energy,
GRB 221009A has brought new insights regarding GRBs
not found so far, such as
the observation of the TeV afterglow \citep{LHAASO2023,LST2025},
non-detection of neutrino \citep{Murase2022,Ai2023,AbbAck2023,VerFra2024},
and so on \citep[see also e.g.,][]{DasRaz2023,ZhaMur2025,KusAsa2025}.

%
%
One of the remarkable discoveries associated with GRB~221009A
is the detection of a mysterious gamma-ray line emission in
${\rm MeV}$ range. 
This spectral feature, identified by \cite{RavSal2024,ZhaXio2024,BurLes2024},
was observed between
246~s and 360~s after the trigger time of {\it Fermi}/GBM.
This line exhibits a time-varying central energy,
$\varepsilon_{\rm line}$,
from $\sim37\,{\rm MeV}$ to $\sim 6\,{\rm MeV}$,
with a relatively narrow spectral width of
$\Delta\varepsilon\sim 2\,{\rm MeV}$.
Electron-positron pair annihilation in the jet
naturally produces such line emissions in the MeV energy range.
This scenario requires high-latitude emission from the GRB jet
to explain the observed time evolution \citep{ZhaLin2024,PeeZha2024}.
Other possible interpretations are also suggested:
A line emission from heavy elements \citep{WeiRen2024};
optical depth discussion \citep{YiZha2024};
down-Comptonization \citep{LiuMao2025};
neutron capture scenario \citep{ZHuFen2025}.
Probable evidence for another MeV line emission has also 
been reported in GRB~221023A \citep{JiaWan2025}.
Such a line emission from GRBs
had already been studied prior to this observation
\citep{PeeWax2004,PeeMes2006,IokMur2007,MurIok2008}.

%
%
The detection of the MeV line from a GRB implies that similar line
features could be observable in high-energy astrophysical
transients beyond GRBs.
Line emissions arising from the electron-positron pair annihilation
may commonly occur in fireballs of MGFs and MSBs.
In particular, MGFs are among the most energetic transient
phenomena observed in the Milky Way Galaxy,
with three events detected to date
(from SGR~0526-66: \citealp{EvaKle1980};
from SGR~1900+14: \citealp{HurCli1999,FerFro1999};
from SGR~1806-20: \citealp{HurBog2005,MerGot2005,BogZog2007,FreGol2007};
see also \citealp{RobVer2021,MerRig2024} for possible extragalactic MGFs).
Motivated by the recent observations of the line emission,
we investigate the spectral properties of electron-positron annihilation 
in MGFs. 
Using a Monte-Carlo approach, we solve the entire sequence of MeV photons
production, propagation, and annihilation, extending previous
one-zone \citep{PeeWax2004,PeeMes2006} and
analytic \citep{IokMur2007,MurIok2008} models developed for line 
emission in GRBs, as well as Monte-Carlo studies for GRBs that omit
pair processes \citep{ItoNag2013,ItoNag2014,Laz2016,ChoLaz2018}.

%
%
This Letter is organized as follows. 
In section~\ref{sec:ana}, we present an analytic model of the
line emission. In section~\ref{sec:num}, we present the results
of 
numerical simulations for the line emission in 
MGFs.
Section~\ref{sec:dis} is devoted to the conclusion and discussion.
Details of our model and numerical simulations can be found in
the Appendices.

\section{Analytic Model of Spectral Shape} \label{sec:ana}
We analytically derive the observed spectrum of line emission
originating from a relativistically expanding fireball.
The method for calculating the analytic spectrum based on numerical simulation is introduced.
For a line emission from a single relativistically expanding
shell,
the observed spectrum is transformed
into a power-law spectrum with cutoffs as a consequence of the 
relativistic Doppler effect. 
The photon index, defined as $dN/d\varepsilon\propto \varepsilon^{-\gamma}$ where $N$ is the photon number and $\varepsilon$ is the photon energy, equals $-1$,
which is determined solely by the geometrical considerations. 
These analytic results will be confirmed by the numerical
calculation in Section~\ref{sec:num}.

To proceed with the analytic derivation, we adopt the
following model assumptions.
We consider a spherically expanding relativistic shell
of radius $r_{\rm em}$ with a half-opening angle $\theta_{\rm j}$,
moving with a Lorentz factor $\Gamma_{\rm em}\,(\gg 1)$.
The luminosity distance between the observer and the shell
is $D_{\rm l}\,(\gg r_{\rm em})$. 
In the comoving frame of the shell,
the photon distribution function is assumed to be isotropic 
\citep[see][for the angular distribution of photons around a photospheric radius]{Bel2011}.
For given fireball dynamics (see Appendix~\ref{sec:app_num} for details), the number of photons originating from pair annihilation is determined around a radius $r_{\rm an}$, where the 
pair-annihilation timescale becomes longer than the dynamical timescale. In contrast, the photon spectrum is determined at the radius $r_{\rm ph}$, where the scattering timescale becomes longer than the dynamical timescale. The former lies inside the latter (see Appendix~\ref{sec:app_num}).
$r_{\rm em}$ lies around $r_{\rm an}$ or $r_{\rm ph}$, and should be determined numerically.

The analytic spectrum is calculated as follows. For photons isotropically distributed in the comoving frame, the photon distribution function is expressed as $f(\varepsilon';\,r_{\rm em})$, where $\varepsilon'$ denotes the photon energy in the comoving frame. Using this distribution function, the observed specific flux is given by (see Appendix~\ref{sec:app_ana} for details)
\begin{eqnarray}
  F_{\varepsilon}&=&
  \frac{2\pi}{c^2}\left(\frac{r_{\rm em}}{D_{\rm l}}\right)^2
        \int_{\mu_{\rm j}}^1 d\mu\,\mu
       \varepsilon^3
       f\left(\varepsilon\Gamma_{\rm em}(1-\beta\mu);\,r_{\rm em}\right),
       \label{eq:specflux}
\end{eqnarray}
where $\beta = (1-\Gamma_{\rm em}^{-2})^{1/2}$, $\mu$ represents the cosine of the angle between the radial direction and the photon momentum, and 
$\mu_{\rm j}$ is the cosine at the edge of the emitting shell.
For a spherical shell, $\mu_{\rm j}$ equals 0.
The photon distribution function at $r_{\rm em}$ is determined by the plasma number density and the pair annihilation rate, which will be calculated numerically using the Monte Carlo method following the procedure of \cite{RamMes1981}.
The analytic spectrum is presented in Section~\ref{sec:num}.

\begin{figure}
\includegraphics[width = 0.5\textwidth]{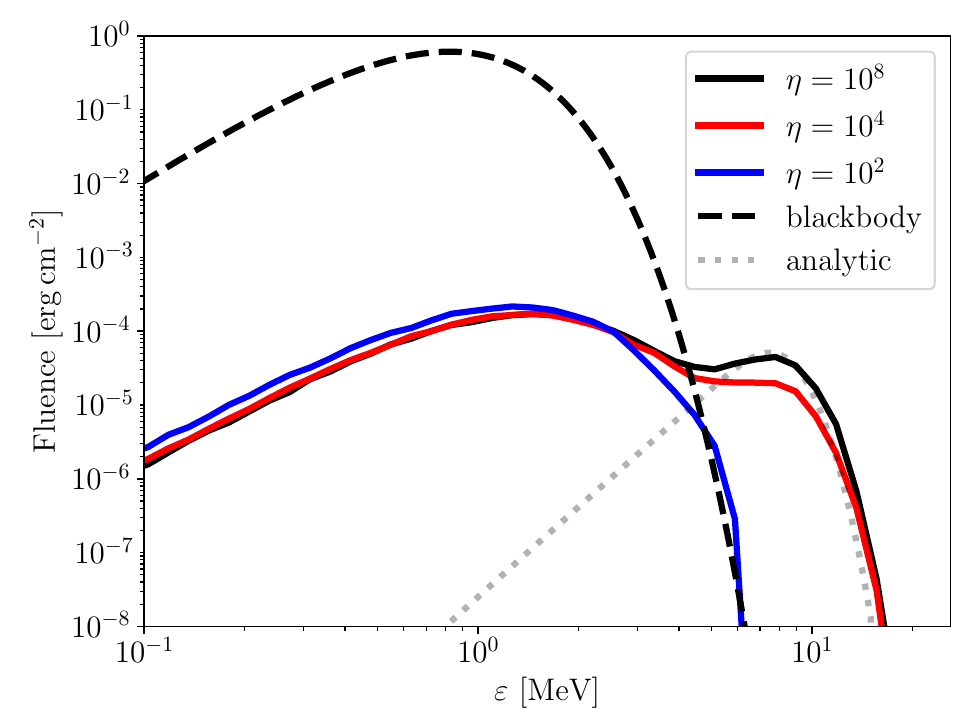}
\caption{Spectra of the observed line emission 
from fireballs with different specific enthalpies (solid lines),
  the primary blackbody radiation (black dashed line),
  and analytic model derived in Section~\ref{sec:ana} (gray dotted line). 
  The adopted parameters are shown in Section~\ref{sec:num}.
  The spectra originally characterized by Equation~(\ref{eq:specflux})
    is modified during propagation 
    due to Compton scattering (see Section~\ref{sec:ana}).
\label{fig:spec}}
\end{figure}

To determine $r_{\rm em}$, where contributes most significantly to the emission spectrum, detailed numerical simulations incorporating photon production, propagation, and absorption are required. We carried out radiation transfer simulations, including these processes, for the parameters of MGFs (see section~\ref{sec:num} for detailed values), and found 
$r_{\rm em}=7.5\times10^6\,{\rm cm}$ (corresponding to the comoving temperature of $T 
\simeq25\,{\rm keV}$, the optical depth for Thomson scattering of $\tau_{\rm T} \simeq 530$).
Using this radius, Equation~(\ref{eq:specflux}), and Monte Carlo method in \cite{RamMes1981}, the photon spectrum can be calculated analytically
(see gray dotted line in Figure~\ref{fig:spec}).

The time dependence of the integral range is ignored, and
we assume that photons emitted from the spherical shell are
observed simultaneously. This assumption is justified as 
follows. For the parameter regime considered in this study, 
the photospheric radius is approximately 
$r_{\rm ph}\sim10^{7}\,{\rm cm}$ and the Lorentz
factor of the photospheric radius is $\Gamma_{\rm ph}\sim 10$. 
Photons emitted from this photospheric radius at a given time
are observed over a timescale of
$t\sim r_{\rm ph}/(c \Gamma_{\rm ph}^2)\sim 10^{-6}\,{\rm s}$,
which would be smaller than the time resolution of
detectors. As a result, photons emitted from $r_{\rm ph}$ at a
given time will fall within the same time bin of a detector
and will be observed simultaneously.
Consequently, it is not necessary to account for the delayed arrival
of high-latitude emission, which is significant in the case of GRBs
\citep{GraPir1999}.

The line-like photon distribution function produces a power-law observed spectrum with a photon index of $-1$. To demonstrate this, we adopt, for simplicity, a Gaussian distribution with mean $\varepsilon'_{\rm line}$ and standard deviation $\varepsilon'_{\rm line}\sigma$, as the distribution function. For this photon distribution function, the resulting spectrum is (see Appendix~\ref{sec:app_ana} for details),
\begin{eqnarray}
F_\varepsilon
 &\propto&\varepsilon^2
 \int_{x_c}^{x_j}dx\left[1-\frac{\varepsilon'_{\rm line}}{\varepsilon\Gamma}(x+1)\right]
 \exp\left[\frac{-x^2}{2\sigma^2}\right],
 \label{eq:specgauss}
\end{eqnarray}
where $x_c = \varepsilon\Gamma(1-\beta)/\varepsilon'_{\rm line}-1$ and
$x_j = \varepsilon\Gamma(1-\beta\mu_{\rm j})/\varepsilon'_{\rm line}-1$.
Because of the Gaussian dependence on $x$ in Equation~(\ref{eq:specgauss}), the dominant contribution to the integral arises at $x\sim 0$.
In this regime, the multiplicative factor preceding the exponential function
in the integrand of Equation~(\ref{eq:specgauss}) becomes
$1-\varepsilon'_{\rm line}/(\varepsilon\Gamma) \sim 1+{\mathcal O}(\Gamma^{-1})$ 
for $\varepsilon\gtrsim \varepsilon'_{\rm line}$.
Therefore, the integral becomes nearly independent of $\varepsilon$
for $\varepsilon\gtrsim \varepsilon'_{\rm line}$, provided that
$x=0$ lies within the integration domain.
In this parameter regime, $F_\varepsilon$ scales as $\varepsilon^2$,
resulting in a power-law spectrum with an index of $2$,
which corresponds to the photon index of $-1$.\footnote{
The specific flux $F_\varepsilon$ is proportional to $\varepsilon (dN/d\varepsilon)\propto \varepsilon^{1-\gamma}$.}
For $\varepsilon$ satisfying $x_c>0$, the flux suddenly decreases since the $x\sim 0$ lies outside the integral range.
This condition gives an upper cut-off energy of $\varepsilon = 2\Gamma \varepsilon'_{\rm line}$.
Therefore, the power-law and cut-off spectrum is formed solely by the geometrical considerations, although it can be modified by further scatterings.

The spectrum of the line emission, which is originally
characterized by $F_\varepsilon\propto \varepsilon^2$,
can be modified during propagation through the surrounding
plasma due to Compton scattering.
Assuming a cold electron population, the optical depth for
Compton scattering is given by \citep{AbrNov1991},
\begin{eqnarray}
  \tau_{\rm C} = \int_{r}^{D_{\rm l}}\frac{dr}{\mu}
  \Gamma(1-\beta\mu)\sigma_{e\gamma}
  n_{\pm}(r),
  \label{eq:tauc}
\end{eqnarray}
where $\sigma_{e\gamma}$ is scattering cross section
for Compton scattering, and $n_{\pm}(r)$ is the electron-positron
number density in the plasma comoving frame.
For $\Theta=\cos^{-1}\mu\ll 1$ and $\Gamma\gg 1$, the factor 
$\Gamma(1-\beta\mu)$ in Equation~(\ref{eq:tauc})
can be expressed as 
$\Gamma^{-1}\left(1+(\Gamma\Theta)^2\right)/2$.
In the regime $\Theta<\Gamma^{-1}$, this reduces to
$\sim \Gamma^{-1}$, substantially  suppressing the scattering.
In contrast, for $\Theta\geq\Gamma^{-1}$, the factor becomes
$\sim \Gamma\Theta^{2}$ increasing with the angle $\Theta$.
As a result, photons originating from high latitudes are
more frequently scattered by the plasma, resulting in the
modification of the initial spectrum. 
This modification 
will be numerically confirmed in Section~\ref{sec:num}.

\section{Numerical Calculation} \label{sec:num}
In this section, we present detailed radiative transfer simulations of 
line emission from
fireballs associated with MGFs.
Using a Monte Carlo method, we model the evolution of the photon field 
under electron-positron pair annihilation, Compton scattering, and pair
production processes. The simulations reveal a characteristic spectrum 
near 10~MeV. We evaluate the detectability of this spectral feature 
with current and future gamma-ray observatories, demonstrating that 
instruments such as the {\it Fermi}/GBM,
e-ASTROGAM,
GRAMS (Gamma-Ray and AntiMatter Survey),
and the All-sky Medium Energy Gamma-ray Observatory eXplorer 
(AMEGO-X) have sufficient
sensitivity to observe the predicted MeV emission from Galactic MGFs.

%
The numerical calculations are performed as follows (see 
Appendix~\ref{sec:app_num} for details). 
The plasma profile is based on analytic solutions for a 
baryon-loaded fireball.
Given the isotropic luminosity $L_{\rm iso}$, the initial fireball 
radius $r_0$, its thickness $\Delta r$, and amount of baryons,
the plasma evolution 
is uniquely determined.
Electron-positron 
pair annihilation (photon injection) is implemented using the Monte-Carlo method
described by \cite{RamMes1981}. At each simulation timestep,
the radial positions of photon injection are randomly sampled 
according to the pair annihilation rate within each cell. 
Once injected, photons propagate outward and either escape the 
computational domain or are annihilated via pair production.
The optical depth and scattering processes are evaluated
following the prescriptions in \cite{AbrNov1991,PozSob1983}.
Electron-positron pair production is treated iteratively 
using the optical depth formula of \cite{GouSch1967}
\citep[see also e.g.,][for details of the iterative procedure]{IwaTak2004}.
The resulting photon distribution function obtained from
each simulation is used in the subsequent simulation to
evaluate the probability of pair production.
This procedure is repeated iteratively until the photon 
spectrum in 2--20~MeV is converged 
(see Appendix~\ref{sec:app_num} for details).
To minimize uncertainties associated with the location 
of the inner boundary, we progressively shifted
the boundary inward
until the spectral results converged.
The fluence of the line emission is hard to evaluate analytically
since it depends on the anisotropic photon field created
by pair annihilation and elementary processes in the anisotropic photon field.
We calculate this fluence numerically using the iterative procedure for
both the photon field and the inner boundary.

%
%
The adopted parameters are an isotropic-equivalent luminosity of 
$L_{\rm iso} = 10^{47}\,{\rm erg\,s^{-1}}$, corresponding to the
brightest MGF observed to date, an initial size of the fireball of
$r_0 = 10^6\, {\rm cm}$, and an observed burst duration of 
$\Delta t = 0.1\,{\rm s}$. The shell width $\Delta r$ is adjusted
to reproduce the observed burst duration. 
For these parameters, 
the initial temperature of the fireball is $190\,{\rm keV}$.
The luminosity distance
to the magnetar is assumed to be $D_{\rm l} = 10\,{\rm kpc}$.
The dimensionless enthalpy $\eta$, which is the radiation energy density 
divided by the baryon rest-mass energy density, is set to be
$10^2$, $10^4$, $10^8$.
For $\eta =10^8$, the baryonic component is negligible in terms of mass and number density, and thus the same as the purely pair fireball. The case with $\eta = 10^2$ is just between the radiation-dominated fireball and baryon-dominated fireball at the photospheric radius. 
The timescale of pair annihilation for electrons is smaller than the dynamical 
timescale in $r<r_{\rm an}=8.2r_0$, and this value is almost independent 
of $\eta$ (see Appendix~\ref{sec:app_num} for details).
The photospheric radii defined by the Thomson cross section are 
at $r_{\rm ph}/r_0 = 8.8,\, 19,\, 86$ for the cases of $\eta = 10^8,\, 10^4,\, 10^2$, 
respectively. At each of these radii, the bulk Lorentz factors are
11, 23, and 105.

%
In Figure~\ref{fig:spec}, 
the black, red, and blue solid line shows the observed gamma-ray fluence
for the cases of $\eta = 10^8$, $10^4$, and $10^2$, respectively.
The black dashed line indicates the primary blackbody component
emitted from the fireball.
The gray dotted line shows the result of the analytic model with $r_{\rm em}/r_0 = 7.5$ (Equation~\ref{eq:specflux}).
For the cases of $\eta = 10^8$, $10^4$, owing
to the relativistic Doppler shift, the peak of the line emission
appears at $\sim 2\Gamma_{\rm em} m_e c^2 \sim 10 \,{\rm MeV}$. 
The power-law spectrum is further modified below $\sim 5\,{\rm MeV}$
due to Compton scattering, as described below Equation~(\ref{eq:tauc}).
This down-scattering softens the spectrum relative to the prediction of
Equation~(\ref{eq:specgauss}), although emission in this energy range
is dominated by the primary blackbody component and cannot be observable.
In MGFs, the emission arising from electron-positron pair annihilation
appears around 10~MeV as an additional component characterized
by a power-law shape with an exponential cutoff, rather than
a pure line feature.

\begin{figure}
\includegraphics[width = 0.5\textwidth]{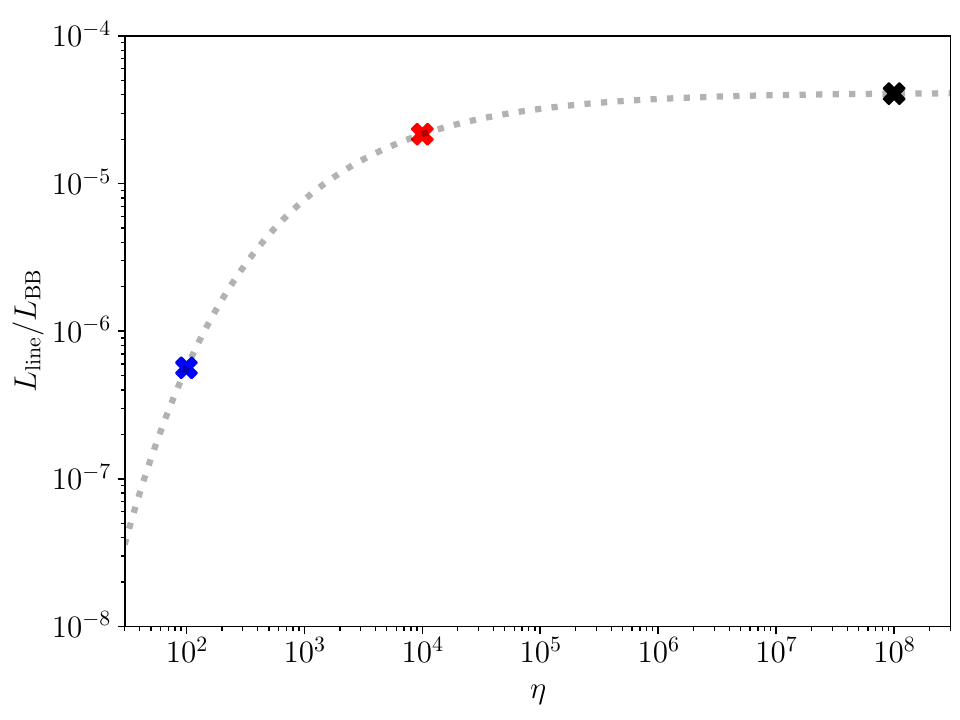}
\caption{$\eta$-dependence of the line luminosity $L_{\rm line}$ normalized 
by the blackbody luminosity $L_{\rm BB}$.
The color scheme follows that of Figure~\ref{fig:spec}
for ease of comparison. \label{fig:ratio}}
\end{figure}
The fewer baryons are present, the brighter the MeV photons become. The reason is as follows. If baryons exist, the associated electrons make the fireball optically thick far beyond $r_{\rm an}$. In this region, the high-energy photons are cooled through Compton scatterings with non-relativistic electrons.\footnote{
For photons with an energy of $m_e c^2$ in the comoving frame, 
the optical depth calculated using the Klein-Nishina cross section at radius $r_{\rm an}$ is 0.5, 8.8, and 880 for the cases of $\eta = 10^8,\,10^4,$ and $10^2$, respectively.}
Since the production rate of high-energy photons does not depend on the specific enthalpy $\eta$ (see Appendix~\ref{sec:app_ana} for details), no additional high-energy photons are generated outside $r_{\rm an}$ in each model. Consequently, the presence of baryons only leads to the cooling of photon energies.
Moreover, the isotropized high-energy photons can be absorbed more efficiently via pair production due to the larger angle between wave vectors.
Figure~\ref{fig:ratio} shows the luminosity ratios of the line component, 
defined as the component brighter than the blackbody above 1 MeV, 
to the blackbody component. 
The dashed line shows a numerical fitting by 
\begin{eqnarray}
    \frac{L_{\rm line}}{L_{\rm BB} }
    = {\mathcal R} \exp\left[-\left(\frac{\eta_c}{\eta}\right)^{s}\right],
\end{eqnarray}
where ${\mathcal R} = 4.1\times10^{-5}$, $\eta_c=3.5\times10^3$, and $s = 0.41$ are adopted.

\nointerlineskip\begin{deluxetable}{cccc}
\tablecaption{Expected photon counts in each energy band of e-ASTROGAM.
\label{tab:photon}}
\tablehead{
\colhead{\shortstack{Energy Band \\(MeV)}} &\colhead{\shortstack{Effective Area\\ (${\rm cm^2}$)}}& \colhead{\shortstack{Photon Counts \\$\eta = 10^8$}} &\colhead{\shortstack{\\$\eta = 10^4$}}
}
\startdata
5.0--15  &  50 &120 & 58\\
7.5--15  & 215 &160 & 71 \\
15--40   & 846 &0.1&0.07
\enddata
\tablecomments{The photons in the first 
row is detected 
via Compton scattering, and those in the last two rows 
via pair production. Effective areas are taken from \cite{eASTROGAM2018}}.
\end{deluxetable}

This line component associated with an initial spike of an MGF could be observable
with current and future gamma-ray telescopes, 
although the detectability depends
on the amount of baryons.
The 
GBM on board the {\it Fermi} satellite is a potential instrument
for detecting this line emission \citep{FermiGBM2009}. The fluence of the
line component shown in Figure~\ref{fig:spec} is comparable to the Ravasio
line reported in GRB 221009A \citep{RavSal2024,ZhaXio2024,BurLes2024}. 
Thus, assuming that the observational instrument is not significantly 
affected by the primary blackbody component, this line component should
be detectable with this existing facility.
More promising instruments are 
e-ASTROGAM \citep{eASTROGAM2017,eASTROGAM2018}, 
GRAMS \citep{GRAMS2020},
and AMEGO-X \citep{AMEGOX2022},
planned space missions
dedicated to MeV-GeV gamma-ray astrophysics.
Table~\ref{tab:photon} shows the  
expected photon counts in each energy band based on 
the effective area of e-ASTROGAM
\citep[Tables 1.3.1 and 1.3.2 in][]{eASTROGAM2018}.
The 1$\sigma$ energy resolution of e-ASTROGAM at 10~MeV is expected to
be less than 0.1~MeV, which will enable the detection of the power-law
spectral shape predicted in this study.
At 10~MeV, the effective areas of AMEGO-X and GRAMS range from  
$20\,{\rm cm^2}$ to $200\,{\rm cm^2}$ (see Figure 4 of
\citealp{GRAMS2020} and Figure 12 of \citealp{AMEGOX2022}). Thus,
both detectors are also capable of detecting ${\mathcal O}(100)$ photons from MGFs.
Accordingly, the detection of an additional component in the MeV range,
originating from the initial spike of a Galactic MGF and characterized 
by a power law plus an exponential cutoff, is expected to yield
approximately ${\mathcal O}(100)$ photons collected within a 0.1~s interval.

\section{Summary and Discussion}\label{sec:dis}
In this paper, we have studied the photon spectrum of electron-positron
pair annihilation line emission in MGFs. 
In section~\ref{sec:ana}, we calculate the observed spectral shape under
the assumptions that photons are emitted from a single shell moving at a
relativistic velocity and that all photons are observed simultaneously.
Through analytic calculations, we have shown 
that the spectrum becomes a power-law distribution with the peak energy
at $2\Gamma\varepsilon'_{\rm line}$.
In section~\ref{sec:num}, we investigated the spectral properties
using Monte Carlo radiative transfer simulations. The calculations
included electron-positron annihilation, Compton scattering, and
pair production. Iteratively evolving the photon field until convergence
enable us to calculate the exact fluence for this line component.
The obtained spectrum is in agreement with the analytic model, with
small modification due to the Compton scattering.

This line emission component from MGFs is potentially detectable
with current instruments such as the {\it Fermi}/GBM, 
although the detectability depends on the specific enthalpy of the fireball.
The proposed MeV gamma-ray satellites, such as e-ASTROGAM, GRAMS, and AMEGO-X
with high energy resolution and sensitivity in the
MeV range, is expected to detect ${\mathcal O}(100)$ photons from such events,
enabling identification of the predicted power-law spectrum with an
exponential cutoff.

This MeV component has not been reported so far. There are several possible
reasons for this. 
First, the most recent Galactic MGF from SGR~1806-20,
which occurred in late 2004, took place before the launch of the {\it Fermi}
satellite in 2008. 
Second, the previous MGF likely involved a large baryon loading,
resulting in weak emission in the MeV range.
Given the limited number of observed MGFs, the possibility of future 
MGFs with small specific enthalpy cannot be excluded. On the contrary, 
MeV observation will independently provide valuable constraints on the 
baryon loading of MGFs,
complementing the X-ray polarization approach for MSBs 
\citep{Wad2025}.
Third, no previous research has focused
on the MeV emission of the initial spike of MGFs. 
Gamma-ray emission from the MGFs in the $0.1$--$10\,{\rm MeV}$ range is reported in \cite{MerGot2005,FreGol2007,BogZog2007}.
However, their analyses targeted the pulsating tail and later phases, rather than the initial spike. 
In these phases, MeV photons may result from  low-energy photons from a trapped fireball being up-scattered \citep[e.g.,][]{YamLyu2020}, or may originate from the decay of r-process elements \citep{PatMet2025}.
Our study underscores the importance of MeV gamma-ray
observation of the initial spike of MGFs.
%

The observed spectrum may exhibit variations due to unconsidered parameters in
this study. Radio afterglow observations of an MGF from SGR~1806-20
\citep{GaeKou2005,CamCha2005} suggest that relativistic outflows associated with
MGFs can contain a baryonic component \citep{NakPir2005}. The presence of this
baryonic component can modify the line spectrum, as MeV photons undergo scattering
by the electrons associated with the baryonic component
\citep{ShePir1990,MesLag1993,MesRee2000}.
This scattering process could lead to a softening of the spectrum, particularly
at lower energies. Furthermore, the isotropic-equivalent luminosity can vary considerably.
Our current analysis focuses on most optimistic case observed to date.
These parameters will be studied in our future studies.

Reproducing the Ravasio line within a simple fireball scenario is unlikely.
To produce a $\sim 10$~MeV line at the maximal isotropic luminosity of 
$L_{\rm iso} = 10^{54}\,{\rm erg\,s^{-1}}$, the initial radius must be 
$r_0 \simeq 10^9\,{\rm cm}$, otherwise the value of Lorentz factor at the 
emission region would be far from the suitable value of $\Gamma_{\rm em}\sim 20$. 
Numerical calculations with this setup yield
a line flux of $L_{\rm iso} \simeq 10^{48}\,{\rm erg\,s^{-1}}$, or 
$\sim 10^{-6}$ times the primary blackbody component.
In contrast, the observed Ravasio line reaches $\sim 10^{-2}$ times the
primary component, ruling out a pure fireball origin.

\begin{acknowledgments}
We thank 
K. Asano, K. Ioka, W. Ishizaki, R. Kuze, R. Matsui, K. Murase, and K. Toma
for fruitful discussions, comments.
We also thank the anonymous referee for many valuable comments that helped improve the manuscript.
This work is supported by Grants-in-Aid for Scientific Research, 
Nos. 22K20366, 23H04899, 25KJ0024, 25K17378 (T.W.), 22K14028, 21H04487, 23H04899 (S.S.K.) from the Ministry of Education, Culture, Sports, Science and Technology (MEXT) of Japan.
S.S.K. acknowledges the support by the Tohoku Initiative for Fostering Global Researchers for Interdisciplinary Sciences (TI-FRIS) of MEXT's Strategic Professional Development Program for Young Researchers.
Discussions during the YITP workshop YITP-W-24-22 on ``Exploring Extreme Transients" 
at the Yukawa Institute for Theoretical Physics in Kyoto University (YITP) were useful
in completing this work. 
The numerical calculations were carried out on Yukawa-21 at YITP and 
Cray XD2000 at the Center for Computational Astrophysics, National Astronomical Observatory of Japan.
\end{acknowledgments}
\appendix
\section{Derivation of Observed Spectrum}\label{sec:app_ana}

\begin{figure}
\includegraphics[width = 0.45\textwidth]{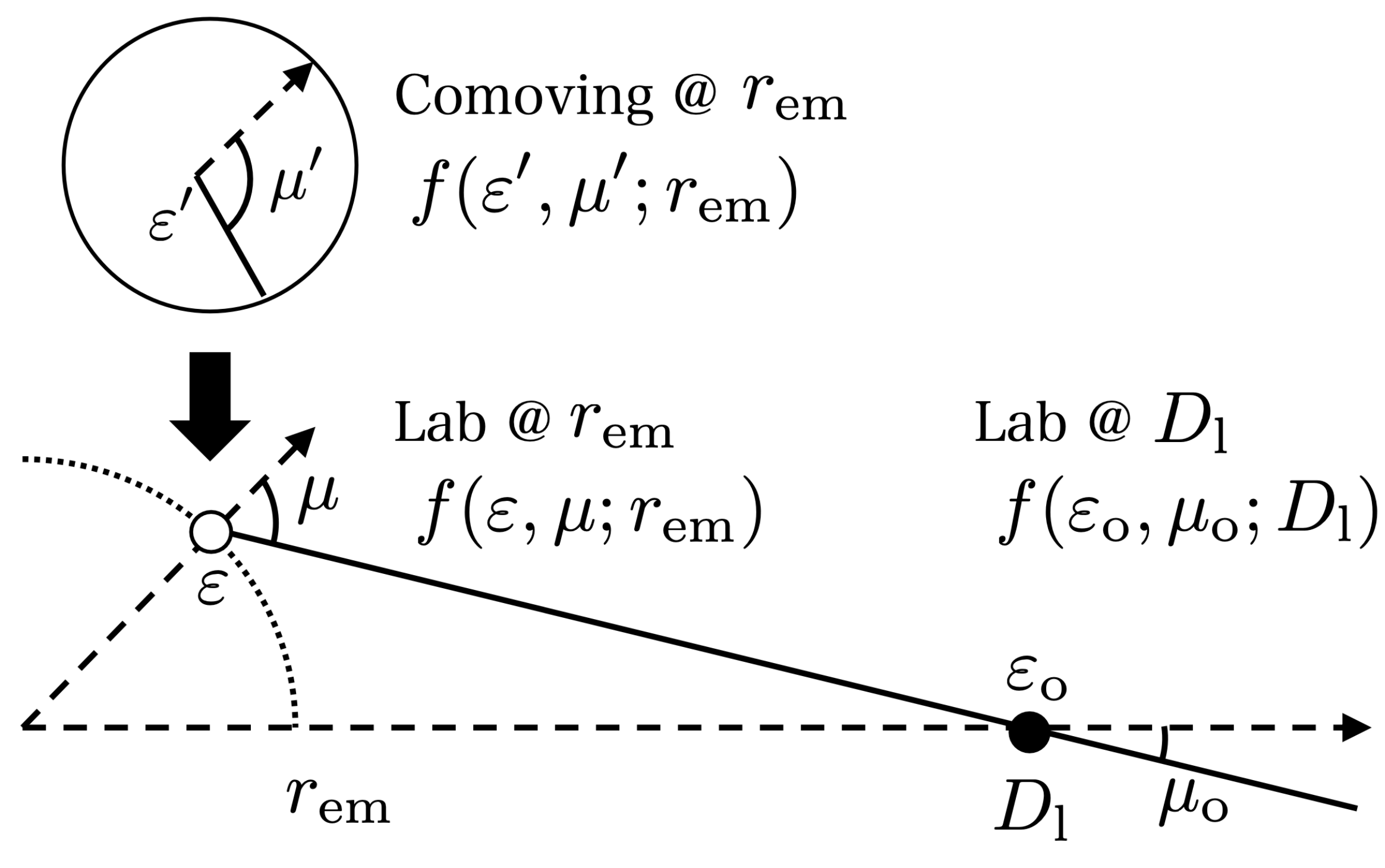}
\caption{Schematic diagram of the variables in Equation~(\ref{eq:appf}).
A photon is emitted at $r_{\rm em}$ isotropically in the plasma comoving frame and is observed at $D_{\rm l}$.
\label{fig:angle}}
\end{figure}

In this section, we derive the relation between the photon distribution 
function and the observed spectrum given in Equation~(\ref{eq:specflux}).
Following the approach in Section~\ref{sec:ana}, we assume that photons 
are emitted from a spherically expanding, relativistic single shell at 
$r_{\rm em}$, with Lorentz factor $\Gamma_{\rm em}$.

For a spherically symmetric system in real space, a photon distribution function
can be expressed as $f(\varepsilon,\mu;r)$,
where $\varepsilon$ is the 
photon energy, $\mu$ is the cosine of the angle between the radial direction
and the photon momentum, and $r$ is the radial coordinate
(see Figure~\ref{fig:angle}).
This distribution function 
is Lorentz invariant and
is related to the specific intensity $I_\varepsilon$,  by 
$I_\varepsilon =\varepsilon^3 f/c^2$ \cite[e.g.,][]{RybickiLightman1979}.
The observer is located at $r=D_{\rm l}$. According to Liouville's theorem, 
the distribution function is conserved along photon trajectories, yielding
\citep{LiSar2008}
\begin{eqnarray}  
f(\varepsilon',\mu';r_{\rm em}) 
=f(\varepsilon,\mu;r_{\rm em})
=f(\varepsilon_{\rm o},\mu_{\rm o};D_{\rm l}),
\label{eq:appf}
\end{eqnarray}
where primed quantities refer to values in the comoving frame of the shell,
and the subscript ‘o’ refers to values in the laboratory frame at the observer position
(see Figure~\ref{fig:angle}).
The first equality in Equation~(\ref{eq:appf})
arises from the Lorentz invariance of the distribution function,
and the second one from Liouville's theorem.
Note that the argument in $f(\varepsilon',\mu';r_{\rm em})$ includes 
variables in both the plasma comoving frame $(\varepsilon',\mu')$
and the laboratory frame $r_{\rm em}$, as is often adopted
\citep[e.g.,][]{LiSar2008,Bel2011}. 

The observed spectrum is a superposition of photons, with
varying beaming factors, over the entire shell.
Using the distribution function, we can calculate the observed specific 
flux as follows.
The observed specific flux is 
\begin{eqnarray}
  F_{\varepsilon_{\rm o}}
  &=&c^{-2}\int d\Omega_{\rm o} \,\varepsilon_{\rm o}^3 f(\varepsilon_{\rm o},\mu_{\rm o};D_{\rm l})\mu_{\rm o}\nonumber\\
  &=&
  \frac{2\pi}{c^2}\left(\frac{r_{\rm em}}{D_{\rm l}}\right)^2
  \int_{\mu_{\rm j}}^1 d\mu\,\mu
  \varepsilon_{\rm o}^3
  f\left(\varepsilon_{\rm o},\mu,;r_{\rm em}\right)
  \label{eq:appFe}       
\end{eqnarray}
where we have used Equation~(\ref{eq:appf}), 
$\mu_{\rm o}^2=1-(r_{\rm em}/D_{\rm l})^2(1-\mu^2)$ \citep[e.g.][]{RybickiLightman1979},
and $\varepsilon_{\rm o}=\varepsilon$. The integral range of $\mu$ is determined 
by the geometry of the emitting shell.
Thus, Equation~(\ref{eq:specflux}) is derived.

Let us assume that the distribution function in the comoving frame is isotropic
and Gaussian, and derive a power-law distribution of observed spectrum
(see Figures 2 and 3 in \citealp{RamMes1981} for the spectral line profile adopted in our simulation).
The mean and standard deviation of the distribution function are set to be $\varepsilon'_{\rm line}$ and $\varepsilon'_{\rm line}\sigma$. The distribution function is expressed as 
\begin{eqnarray}
f_{\rm G}(\varepsilon')=\frac{{\mathcal F}_0}{\sqrt{2\pi\sigma^2}}\exp\left[-\frac{(\varepsilon'/\varepsilon'_{\rm line}-1)^2}{2\sigma^2}\right],
\label{eq:Fgauss}
\end{eqnarray}
where ${\mathcal F}_0$ is a constant representing the photon number
density. For $\sigma\ll 1$, the photon number density in
real space is given by $4\pi c^{-3}\varepsilon'^3_{\rm line}{\mathcal F}_0$.
For an isotropic distribution
in the comoving frame, $f(\varepsilon',\mu';r_{\rm em})$ is independent of $\mu'$ 
and can be expressed as 
$f_{\rm G}(\varepsilon')$ at $r = r_{\rm em}$.
Using Equation~(\ref{eq:appf}) together with the relation
$\varepsilon'=\varepsilon_{\rm o}\Gamma(1-\beta\mu)$,
Equation~(\ref{eq:appFe}) can be rewritten as
\begin{eqnarray}
  F_{\varepsilon_o}&=&
  \frac{2\pi}{c^2}\left(\frac{r_{\rm em}}{D_{\rm l}}\right)^2
        \int_{\mu_{\rm j}}^1 d\mu\,\mu
       \varepsilon_o^3
     f_{\rm G}\left(\varepsilon_o\Gamma(1-\beta\mu)\right).
\end{eqnarray}
Substituting a gaussian distribution (see Equation~\ref{eq:Fgauss}) into this
expression, we obtain 
the observed spectrum as follows,
\begin{eqnarray}
F_{\varepsilon_{\rm o}}&=&\frac{2\pi}{c^2}\left(\frac{r_{\rm em}}{D_{\rm l}}\right)^2
        \int^1_{\mu_{\rm j}} d\mu\,\mu
       \varepsilon_{\rm o}^3
       \frac{{\mathcal F}_0}{\sqrt{2\pi \sigma^2}}\nonumber\\
        & & \times    \exp\left[-\frac{\left(\varepsilon_{\rm o}\Gamma(1-\beta\mu)/\varepsilon'_{\rm line}-1\right)^2}{2\sigma^2}\right]\nonumber\\
    &=&\frac{\sqrt{2\pi}}{c^2\sigma}\left(\frac{r_{\rm em}}{D_{\rm l}}\right)^2\frac{\varepsilon'_{\rm line}}{\Gamma\beta^2}\varepsilon_{\rm o}^2 \mathcal{F}_0\nonumber\\
& &\times\int_{x_{\rm c}}^{x_{\rm j}}dx\left[1-\frac{\varepsilon'_{\rm line}}{\varepsilon_{\rm o}\Gamma}(x+1)\right]\exp\left[\frac{-x^2}{2\sigma^2}\right],\nonumber\\
\end{eqnarray}
where the integral variable $\mu$ is transformed to $x$ using the relation
$x=\varepsilon_{\rm o}\Gamma(1-\beta\mu)/\varepsilon'_{\rm line}-1$.
The integral range for $x$ is from
$x_{\rm c}=\varepsilon_{\rm o}\Gamma(1-\beta)/\varepsilon'_{\rm line}-1$
to 
$x_{\rm j}=\varepsilon_{\rm o}\Gamma(1-\beta\mu_{\rm j})/\varepsilon'_{\rm line}-1$.
Rewriting $\varepsilon_{\rm o}$ as $\varepsilon$ and substituting $\beta = 1$, 
we obtain Equation~(\ref{eq:specgauss}).

The spectral shape is determined by the order of magnitude of $x_c$, 0, and $x_j$.
The integral domain is
\begin{eqnarray}
\frac{\varepsilon\Gamma(1-\beta)}{\varepsilon'_{\rm line}}-1
\leq x
\leq \frac{\varepsilon\Gamma(1-\beta\mu_{\rm j})}{\varepsilon'_{\rm line}}-1,
\label{eq:dx}
\end{eqnarray}
which is determined by the angular size of the shell
$1\geq \mu \geq \mu_{\rm j}.$
The dominant contribution to the integral arises around $x\sim 0$.
As mentioned in Section~\ref{sec:ana}, if $x_c\lesssim 0\lesssim x_j$ is satisfied,
the integral becomes nearly independent of $\varepsilon$ resulting in  $F_\varepsilon\propto \varepsilon^2$.
At lower energies, the spectrum begins to slightly deviate
from this power-law around $\varepsilon\sim \varepsilon'_{\rm line}$,
and exhibits a sudden decline at the energy where $0>x_j$ is satisfied.
The upper cutoff energy appears at the energy where $0<x_c$ is satisfied.
Therefore, cutoff occurs outside the following energy range:
\begin{eqnarray}
  \frac{\varepsilon'_{\rm line}}{\Gamma(1-\beta\mu_{\rm j})}
  \leq \varepsilon \lesssim
  2\Gamma\varepsilon'_{\rm line},
 \label{eq:range}
\end{eqnarray}
the spectrum 
where we have used $(1-\beta)^{-1}\simeq 2\Gamma^2$ for $\Gamma \gg 1$.
Within this energy range, the power-law spectrum is formed.
For $\varepsilon'_{\rm line} = 511\,{\rm keV}$, $\Gamma = 10$, and $\mu_{\rm j} =0$, the upper cutoff is approximately 10 MeV, and the lower cutoff is about 50 keV.

%
%
\section{Details of Numerical Calculation}\label{sec:app_num}
We solve the radiative transfer using a  Monte-Carlo scheme applied to 
the plasma flow described by the analytic solution of relativistic fireballs.
In this section, we describe the analytic solution for a 
baryon-loaded fireball,
which is adopted in this study, and describes the details of the numerical
method.
In this paper, we focus on the parameter regime where the radiation energy density at the photospheric radius exceeds the baryon rest-mass energy density, as this condition is required to produce the blackbody radiation observed in MGFs.

\begin{figure}
\includegraphics[width = 0.47\textwidth]{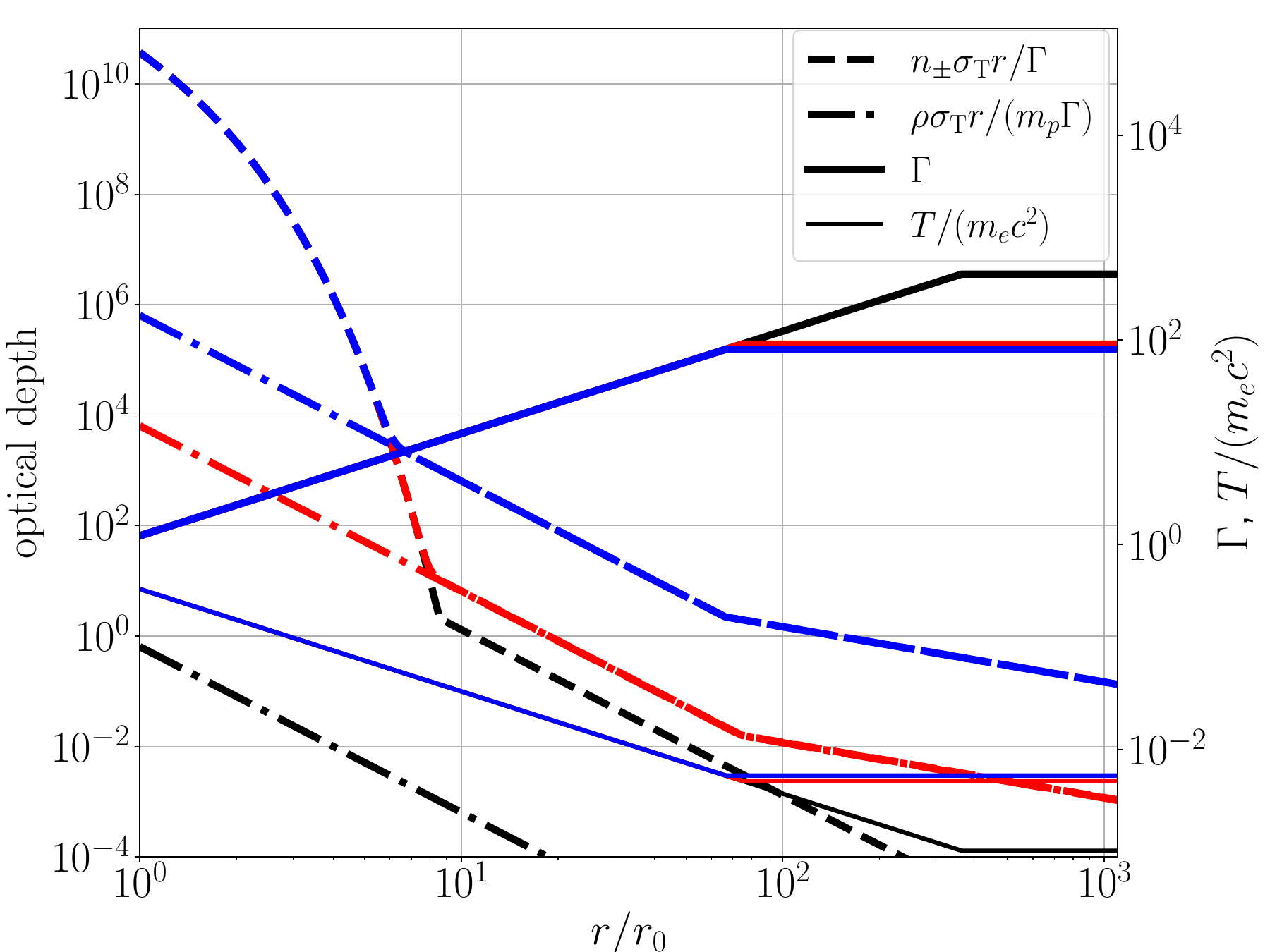}
\caption{Radial dependences of optical depth for Thomson scattering 
(dashed lines for the total electron-positron population and dash-dotted lines for electrons associated with the baryonic component), Lorentz factor (thick solid lines), and comoving temperature (thin solid lines). 
The color scheme follows that of Figure~\ref{fig:spec}, i.e., black, red, and blue lines correspond to the cases with $\eta = 10^8$, $10^4$, and $10^2$, respectively.
\label{fig:radius}}
\end{figure}

The analytic solution for a 
baryon-loaded fireball is summarized 
below (see \citealp{Pac1986,ShePir1990,MesLag1993,MesRee2000}; 
see e.g., \citealp{Pir1999,Zha2018G} for reviews).
The fireball dynamics are determined by the isotropic luminosity, $L_{\rm iso}$,
the initial radius, $r_0$, 
initial baryon-mass density, $\rho_0$,
and the width of a shell, $\Delta r$. 
To describe the dynamics and thermodynamics of an optically thick 
fireball,
we adopt a one-component fluid approximation under the assumption that the plasma
and radiation are in thermal equilibrium initially.
For a given isotropic-equivalent luminosity $L_{\rm iso}$, initial 
radius $r_0$, and initial Lorentz factor $\Gamma_0$, the initial temperature of the fireball is determined as 
\begin{eqnarray}
T_0=\left(\frac{L_{\rm iso}}{4\pi r_0^2 c a_{\rm rad}\Gamma_0^2}\right)^{1/4}
\sim 190\,{\rm keV}\,L_{\rm iso,47}^{1/4},
\end{eqnarray}
where $a_{\rm rad}$ is the radiation constant, and $T_0$ has the dimension 
of energy. 
Assuming that the transonic point lies sufficiently deep inside, 
the velocity at $r_0$ is taken to be the sound speed of a radiation-dominated gas,
$c/\sqrt{3}$, 
resulting in $\Gamma_0=\sqrt{3/2}$.
The dimensionless enthalpy $\eta$ is defined as
\begin{eqnarray}
\eta = \frac{a_{\rm rad}T_0^4}{\rho_0c^2}.
\end{eqnarray}
In the ultrarelativistic limit during the acceleration phase (see e.g., \citealp{Pir1999, Zha2018G} for the coasting phase),
the fireball's bulk Lorentz factor $\Gamma$, comoving temperature $T$,
baryon-mass density $\rho$,
and 
positron number density $n_+$
evolve as 
(e.g., \citealp{Pac1986,Pir1999,Zha2018G}, see also
Section 105 of \citealp{Landau1980stat} for Equation~\ref{eq:app_npm})
\begin{eqnarray}
\Gamma&=&\Gamma_0\bar{r},\label{eq:app_gamma}\\
T&=& T_0 \bar{r}^{-1},\label{eq:app_t}\\
\rho &=& \rho_0\bar{r}^{-3},\label{eq:app_rho}\\
n_+&=&
- \frac{\rho}{2m_{p}}+
  \left[\frac{\rho^2}{4m_{p}^2}+4\left(\frac{m_eT}{2\pi \hbar^2}\right)^{3}
    \mathrm{e}^{-2m_ec^2/T}\right]^{1/2},\label{eq:app_npm}\nonumber\\
\end{eqnarray}
where $\hbar$ is the reduced Planck constant 
and $\bar{r}=r/r_0$. 
From charge neutrality, the electron number density $n_-$ is given by $n_- = n_++\rho/m_p$.
The radial dependences of these values are shown in Figure~\ref{fig:radius}.
As indicated by Equation~(\ref{eq:app_npm}), the positron number density exhibits a complex dependence on $\bar{r}$. However, in the limit where the baryon number density is negligible compared to that of the positrons, the relation simplifies to $n_+\propto \bar{r}^{-3/2}\exp(-m_ec^2T_0^{-1}\bar{r})$, depending on the radius exponentially.
We note that the product $n_+n_-$, which determines the
  pair-annihilation rate, does not depend on the baryon number density $\rho/m_p$.
  This is because the electron and positron chemical potentials $\mu_-,\,\mu_+$
  satisfy $\mu_+ +\mu_-= 0$, and the factor involving the chemical potentials in 
  the number densities, 
  $n_+n_-\propto \exp[(\mu_++\mu_-)/T]$, cancel out.

The photons escape from the radius where the optical depth for Thomson scattering is approximately unity. For electrons associated with the baryonic component, the photospheric radius is defined by the condition $\sigma_T (\rho/m_p) r/\Gamma = 1$, which yields a photospheric radius for baryonic component,
\begin{eqnarray}
    \bar{r}_{\rm ph,baryon} = \left(\frac{\sigma_{T}L_{\rm iso}}{4\pi m_pr_0c^3\Gamma_0^3\eta}\right)^{1/3}
    \sim 86L_{\rm iso,47}^{1/3}\eta_{,2}^{-1/3}.\nonumber\\
\end{eqnarray}
Similarly, for the pair component, the photospheric radius is defined by $\sigma_T 2n_+ r/\Gamma = 1$. In the regime where the baryon number density is negligible compared to the pair number density, $n_+$ is dominated by an exponential dependence on $T$ (see Equation~\ref{eq:app_npm}); consequently the temperature at the photospheric radius depends weakly on other parameters. 
Defining the temperature at this radius as $T_{\rm ph,pair}$ and neglecting the logarithmic paramaeter dependence in the pair number density (see Equation~\ref{eq:app_npm}), the photospheric radius for pair component is given by
\begin{eqnarray}
    \bar{r}_{\rm ph,pair} \simeq \frac{T_0}{T_{\rm ph,pair}}\sim 
    8.8 L_{\rm iso,47}^{1/4},
    \label{eq:app_rphp}
\end{eqnarray}
where we adopt a fiducial value of $T_{\rm ph,pair}=22\,{\rm keV}$, which has a weak logarithmic dependence on $L_{\rm iso}$, $r_0$, and $\eta$.
  
Beyond the annihilation radius $r_{\rm an}$, 
where the timescale of pair annihilation for positrons becomes comparable to the dynamical timescale,
the positron number density freezes out, and the single-component approximation breaks down.
This radius is determined by the condition $(\sigma_{+-} c\beta_e n_-)^{-1} \simeq r/(c\Gamma)$, where $\beta_e$ is the thermal velocity of the pairs normalized by the speed of light. In the non-relativistic limit ($\beta_e\ll1$), the pair-annihilation cross section is approximated by $\sigma_{+-}\simeq3\sigma_T/(8\beta_e)$.
Assuming that the baryon number density is negligible compared to the pair number density, same as in Equation~(\ref{eq:app_rphp}), this radius is mainly determined by the temperature of the plasma, and the annihilation radius is given by
\begin{eqnarray}
    r_{\rm an} \simeq\frac{T_0}{T_{\rm an,pair}}\sim 8.2L_{\rm iso,47}^{1/4},
\end{eqnarray}
where $T_{\rm an,pair}=23\,{\rm keV}$, which also has a weak logarithmic dependence on parameters, is adopted.
In this parameter regime, Equation~(\ref{eq:app_npm}) is no longer valid.
Beyond this radius, the 
positron
number density is determined by the conservation 
of number flux:
\begin{eqnarray}
    n_{+}=n_{+,\rm an}\frac{\Gamma_{\rm an}r_{\rm an}^2}{\Gamma r^2},
    \label{eq:app_npm_f}
\end{eqnarray}
where subscript `an' denotes the value at $r_{\rm an}$.

\begin{figure}
\includegraphics[width = 0.45\textwidth]{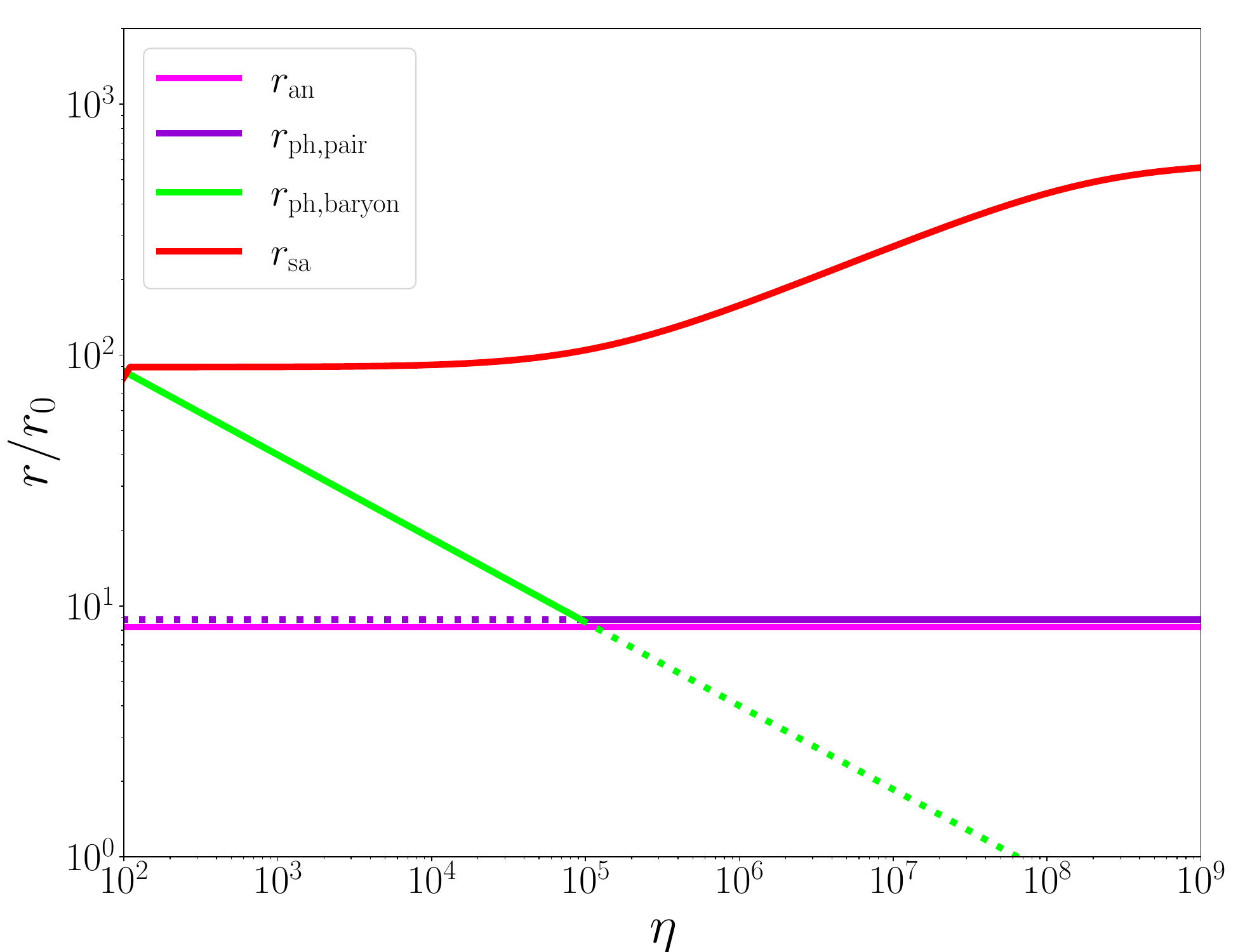}
\caption{$\eta$-dependence of each radius.
$r_{\rm ph,pair}$ and $r_{\rm ph,baryon}$ are photospheric radii for pair-plasma and baryonic component, respectively.
The physically relevant photospheric radius $r_{\rm ph}\simeq \max(r_{\rm ph,pair},r_{\rm ph,baryon})$, is shown with solid green and purple lines; dotted lines show analytic expressions in regimes where the corresponding radii are not physically relevant.
\label{fig:eta}}
\end{figure}

Above the photospheric radius, the plasma continues to be accelerated by scattering of the escaping photons.
Photons decouple from the fireball at the photospheric radius $r_{\rm ph}$, which is defined using the total electron positron number density $n_\pm = n_-+n_+$ and well approximated by $r_{\rm ph}\simeq \max(r_{\rm ph,pair},r_{\rm ph,baryon})$,
but the scaling relations in Equations~(\ref{eq:app_gamma})
still hold.
The acceleration above $r_{\rm ph}$ occurs because the pair plasma is accelerated 
via scattering by the escaping photons, which dominate the energy density,
as long as the photospheric luminosity exceeds the kinetic luminosity of the plasma
\citep{GriWas1998,LiSar2008,ChoLaz2018}.
Above $r_{\rm ph}$, the electron velocity distribution follow the 
Maxwell-Boltzmann distribution, normalized by the number density of
Equation~(\ref{eq:app_npm_f}). The acceleration by the radiation ceases
at the saturation radius,
where the work done by the radiation during the dynamical time in the comoving 
frame is equal to the rest mass energy of a particle
\citep{MesLag1993,GriWas1998,NakPir2005},
\begin{eqnarray}
\bar{r}_{\rm sa}=\left(\frac{L_{\rm iso}\sigma_T}{4\pi \bar{m}c^3 r_0\Gamma_0^3}\right)^{1/4}
\sim 90\,L_{\rm iso,47}^{1/4}\left(\frac{m_p}{\bar{m}}\right)^{1/4},\nonumber\\
\end{eqnarray}
where $\bar{m} = (\rho+m_en_-+m_en_+)/(\rho/m_p+n_-+n_+)$ is the average particle mass.
Shell spreading 
begins at $r\sim \bar{r}_{\rm sa}^2\Delta r $, but this radius lies far beyond the region 
of interest in this study.
$\eta$-dependence of these radii are summerized in Figure~\ref{fig:eta}.

On this analytic plasma profile, we solved the photon propagation numerically.
In our Monte-Carlo scheme, the radiation field is represented by a collection of
photon packets, each carrying information on position, momentum, and number of 
photons in the packet \citep[see e.g.,][]{DolGam2009,Laz2016,KawFuj2023}. 
The radical direction is discretized into $N_r = 4000$ spatial cells on a 
logarithmic scale. Within each cell, the analytic solutions for $T,\,n_\pm$,
$\rho$,
and $\Gamma$, along with the local mean free path for photon pair production
are provided. The optical depth for photon-photon annihilation depends on both 
the photon frequency and its propagation direction. The frequency is discretized
logarithmically over the range $10^{-3}m_ec^2$ to $10^{3}m_ec^2$ with 10000 bins,
while the propagation direction is discretized uniformly in cosine, also with 
10000 bins. For the optical depth calculation of pair production,
both dimensions are coarsened into 100 cells each, and the values at midpoints
are obtained via interpolation. Using the values, the optical depths for the 
scattering and pair production are evaluated at each time step.

We incorporate both the time dependence and photon annihilation via 
electron-positron pair production into the numerical simulation code developed in 
\cite{WadAsa2025}. The code has been validated against several benchmark problems: 
it reproduces the inverse-Comptonized spectrum in a static plasma as presented in \cite{PozSob1983};
the relativistic beaming factor in an outflow as shown in \cite{Bel2011};
the analytic solution for the radiation energy-momentum tensor in a relativistic outflow;
and the analytic optical depth for pair production for the static, power-law photon distribution.
The details of the scattering process are described in \cite{WadAsa2025}. The stochastic 
treatment of pair production and Compton scattering follows the prescription of \cite{IwaTak2004},
and the pair annihilation is incorporated via iterative procedures similar to those 
described therein.
The location of the inner boundary is progressively shifted
inward, starting from a radius corresponding to an optical
depth for Thomson scattering of $\tau_T = 15$ until the 
spectral results converged around $\tau_T\simeq 4000$.
We confirm that the numerical results converged within 10\% accuracy
in the energy range of $2$--$20$~MeV.

\begin{figure}[th]
\includegraphics[width = 0.47\textwidth]{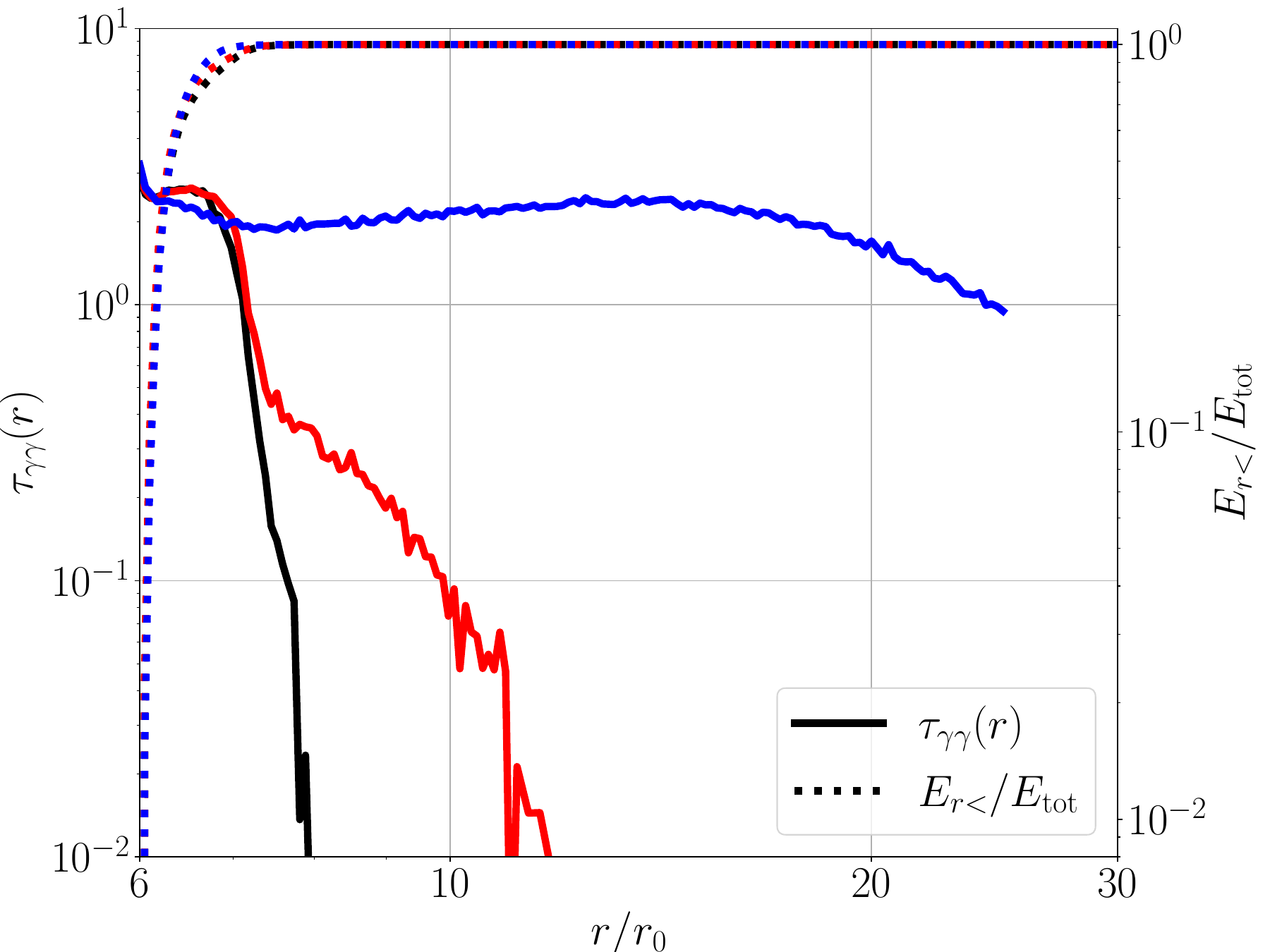}
\caption{Radial dependence of pair-annihilation optical depth for photons emitted at the radius $r$ (solid lines), and radial dependence of enclosed energy fraction (dotted lines). The color scheme follows those of Figures~\ref{fig:spec} and \ref{fig:radius}.
\label{fig:tau}}
\end{figure}

The solid lines in Figure~\ref{fig:tau} show the pair-annihilation optical depth for photons emitted at radius $r$, as evaluated from the simulation results. Photons with $\varepsilon\geq m_ec^2$ are taken into account here. The energy of photons emitted at radius $r$ is attenuated by a factor of $\exp(-\tau_{\gamma\gamma})$ at the outer edge of the simulation domain due to the pair annihilation. Pair-annihilation effects are more pronounced in baryon-rich fireballs.
The optical depth shown by the solid curve in Figure~\ref{fig:tau} is computed by dividing the radius into bins. For each bin, we define $E_{\rm inj}(r)$ as the total energy of photons injected in the bin that would be observed with energies above $m_e c^2$ if they escaped freely, and $E_{\rm out}(r)$ as the total energy of photons from the same region that are actually observed above $m_e c^2$. The optical depth for pair-annihilation is then evaluated by $\tau_{\gamma\gamma}(r) = -\ln\left[E_{\rm out}(r)/E_{\rm inj}(r)\right]$.
The dotted lines in Figure~\ref{fig:tau} show the radial dependence of the enclosed luminosity fraction, defined as the cumulative fraction of total radiated energy originating from within radius $r$. This quantity indicates the contribution of emission generated inside the radius $r$, $E_{<r}$, to the final energy output $E_{\rm tot}$. For the $\eta = 10^2$ case (blue lines), although $\tau_{\gamma\gamma}$ is of order unity at $r\sim 20$ (blue solid line), the energy fraction originating from this radius is negligible (blue dotted line) and does not affect the observed spectrum.

%
\bibliography{cite}{}
\bibliographystyle{aasjournal}



\end{document}